# PRAGMATIC SELF-ADJOINT PROCEDURE IN THE SCHRODINGER EQUATION FOR THE INVERSE SQUARE POTENTIAL

Anzor Khelashvili [*] , Teimuraz Nadareishvili[*,**]

[*] Academy Member, Inst. of High Energy Physics, Iv. Javakhishvili Tbilisi State University, Tbilisi, Georgia.
[**] Department of Physics, Faculty of Exact and Natural Sciences, , Iv. Javakhishvili Tbilisi State University,Tbilisi, Georgia.
E-mail : anzor.khelashvili@tsu.ge ;  teimuraz.nadareishvili@tsu.ge

***ABSTRACT:***

*The self-adjoint extension (SAE) procedure is considered in the Schrodinger equation for potentials behaving as an attractive inverse square at the origin of coordinates. This approach guarantees self-adjointness of the radial Hamiltonian in three dimensions. It is shown that the single bound state appears after such an extension, which depends on SAE parameter. The same parameter arises for the scattering case as well, when the extension is made by orthogonality requirement. The closed form is derived for the modified scattering amplitude, which consists an extra factor depended on the SAE parameter. That guarantees the appearance of the same bound state in the form of the scattering amplitude pole. So, the generalization of pragmatic method is demonstrated in case of continuous spectrum.*

## 1. Introduction

The inverse squared $r^{-2}$ potential receives a widespread attention in various problems of quantum mechanics. Number of physically significant quantum mechanical problems manifest in such a behavior.

Detailed consideration of papers devoted to problems concerning with this potential put in doubt the motivations for neglecting of so-called additional (singular) solutions, which are based on mathematical sets of quantum mechanics without invoking of specific physical ideas.

The aim of this article is to include the singular solution into consideration by using a well-known procedure of self-adjoint extension (SAE), in particular, a "pragmatic approach", which is much simpler than general Weil method.

This article is constructed as follows: Sec I is devoted to discussion of Schrodinger radial equation and applied notations, classification of singular potentials, characterization of accepted areas of parameters. In Sec II the orthogonality requirement is considered for bound states, as the main method of used self-adjoint extension. It is shown, that the hermiticity of Hamiltonian is proportional to orthogonality condition of its solutions.  Therefore, the requirement of orthogonality at the origin guarantees SAE, though this procedure introduces an extra arbitrary parameter. Then in Sec. IV SAE for a pure inverse squire potential is carried out and the existence of a single bound state is established. Further, the Sec V and VI are devoted to SAE in case of scattering (continuous states). The pragmatic orthogonality extension is applied again and the standard, as well as additional (singular) solutions are retained and the appearance of a single bound state, as a pole of partial scattering amplitude is obtained. The corresponding scattering phases are also calculated.

**Bound states (discrete spectrum)**

The full radial wave function $R(r)$ satisfies the following Schrodinger equation in 3-dimensions

$$\frac{d^2R}{dr^2} + \frac{2}{r}\frac{dR}{dr} + 2m[E - V(r)]R - \frac{l(l+1)}{r^2}R = 0 \tag{2.1}$$

The traditional change of variables in quantum mechanics eliminates the first derivative term from this equation by the following substitution

$$R(r) = \frac{u(r)}{r}, \tag{2.2}$$

which in turn, gives the equation for the reduced radial wave function $u(r)$

$$\frac{d^2u(r)}{dr^2} - \frac{l(l+1)}{r^2}u(r) + 2m[E - V(r)]u(r) = 0, \tag{2.3}$$

only if the following boundary (Dirichlet) condition is required

$$u(0) = 0, \tag{2.4}$$

irrespective the potential is regular or singular. The following classification for potentials is known in case of Schrodinger equation.

1. **Regular potentials**: if
$$\lim_{r \to 0} r^2 V(r) = 0 \tag{2.5}$$

    In this case the radial function $u(r)$ behaves as

$$\lim_{r \to 0} u = C_1 r^l + C_2 r^{-(l+1)} \tag{2.6}$$

where $l$ is orbital momentum. The second term in this expression is singular, it does not satisfy the condition (2.4) and therefore, must be neglected $(C_2 = 0)$.

2. **Strong singular potential**, for which

$$\lim_{r \to 0} r^2 V(r) \to \pm\infty \tag{2.7}$$

In this case the "falling onto center" takes place.

3. It is interesting to study potentials with intermediate behavior, called "transitive potentials"
$$\lim_{r \to 0} r^2 V(r) \to \pm V_0 \quad (V_0 = const > 0) \tag{2.8}$$

Two signs here correspond to repulsive (+) and attractive (-) potentials, correspondingly. For such a potential it follows the following asymptotics near the origin

$$\lim_{r \to 0} R(r) = Ar^{-1/2+P} + Br^{-1/2-P} = a_{st} r^{-1/2+P} + a_{add} r^{-1/2-P} \equiv R_{st} + R_{add}, \tag{2.9}$$

where

$$P = \sqrt{(l+1/2)^2 - 2mV_0} > 0 \tag{2.10}$$

Hence we have two regions for this parameter $P$:

$$0 \leq P < 1/2 \quad \text{and} \quad P \geq 1/2 \tag{2.11}$$

In both of them the wave function $R(r)$ is normalizeble

$$\int_0^\infty r^2 dr |R(r)|^2 = 1 \quad \text{(for bound states)} \tag{2.12}$$

In the first interval the second term, $R_{add} = a_{add} r^{-1/2-P}$ must also be retained, because the boundary condition is also fulfilled for it. The potential like (2.8) was firstly considered by K.Case [1], but he ignored the second term in a solution. As regards of a region $P \geq \frac{1}{2}$, only the first term $R_{st} = a_{st} r^{-1/2+P}$ survives.

From the positivity of $P$ and from inequality $P < 1/2$ it follows conditions of existency additional states, which restrict the parameter $2mV_0$ in the following interval

$$l(l+1) < 2mV_0 < l(l+1) + 1/4 \tag{2.13}$$

Intervals from the left and from the right sides have no crossing and therefore, if additional solution exists for fixed $V_0$ and for some $l$, then it is absent for another $l$.

We see that in the $l = 0$ state except the standard solutions there are additional solutions as well for arbitrary small $V_0$, while for $l \neq 0$ the "strong" field is necessary in order to fulfil (2.13).

As was shown in [2] there are no satisfactory arguments in the framework of quantum mechanics, which avoid this additional solution self-consistently.

Therefore, one has to retain this additional solution and study its consequences.

**2. Introduction of the SAE parameter (discrete case)**

The operator $A$ is called Hermitian (or symmetric) if for any functions $u$ and $\hat{}$ it satisfies the condition

$$\langle \hat{} | \hat{A} | u \rangle = \langle \hat{A}^{+\hat{}} | u \rangle = \langle \hat{A}\hat{} | u \rangle \tag{3.1}$$

For self-adjointness it is required in addition that the domains of functions of operators $A$ and $A^+$ would be equal. As a rule, the domain of $A^+$ is wider and it becomes necessary to make a self-adjoint extention of the operator $A$.

There exists a well-known powerful mathematical apparatus for this purposes [3,4]. We use below fairly simple approach, so-called "pragmatic approach", which is much simpler and gets the same results as the strong mathematical full SAE procedure. Moreover, this method is physically more transparent.

Considering the full radial Hamiltonian

$$H_R = -\frac{d^2}{dr^2} - \frac{2}{r}\frac{d}{dr} + \frac{l(l+1)}{r^2} + 2mV(r), \tag{3.2}$$

it is easy to see that for any two eigenfunctions $R_1$ and $R_2$ corresponding to the levels $E_1$ and $E_2$ of it, the condition (3.2) takes the following form

$$\int_0^\infty R^*_1 \hat{H}_R R_2 r^2 dr - \int_0^\infty R^*_2 \hat{H}_R R_1 r^2 dr = 2m(E_2 - E_1)\int_0^\infty R^*_1 R_2 dr \quad (3.3)$$

We see that a self-adjointness condition is proportional to the orthogonality integral, therefore these two conditions are mutually dependent. As the self-adjoint operator has orthogonal eigenfunctions, requirement of orthogonality automatically provides self-adjointness of $H_R$, which means that this way provides a realization of SAE procedure. It is an essence of the "pragmatic approach"[5].

In case of regular potentials (2.5), as mentioned above, we retain only regular solutions, which behaves as

$$R_{st} \underset{r\to 0}{\sim} a_{st} r^{l+1} \quad (3.4)$$

Calculating the right-hand side of Eq.(3.3) by this behavior, and taking into account the decaying asymptotic at infinity (bound states), we get zero. Therefore, for regular potentials the radial Hamiltonian $H_R$ is a self-adjoint operator.

Contrary to this, for the inverse square potential one has to retain the additional solution $R_{add} \underset{r\to 0}{\sim} a_{add} r^{-1/2-P}$ as well. Now the right-hand side of (3.1) is not zero in general, but is

$$m(E_1 - E_2)\int_0^\infty R^*_1 R_2 r^2 dr = P\{a_{1st} a_{2add} - a_{1add} a_{2st}\} \quad (3.5)$$

where constants $a_{ist}, a_{iadd}$ $(i=1,2)$ are defined in (2.9).

Thus, retaining of additional solution causes breakdown of orthogonality and, consequently, $H_R$ is no more a self-adjoint operator.

It is natural to ask – how to fulfil the orthogonality condition? It's clear, one must require

$$a_{1st} a_{2add} - a_{1add} a_{2st} = 0 \quad (3.6)$$

or equivalently

$$\frac{a_{1add}}{a_{1st}} = \frac{a_{2add}}{a_{2st}} \quad (3.7)$$

In this case the radial Hamiltonian $H_R$ becomes a self-adjoint operator. This generalizes the Case result [3].

Therefore it is necessary to introduce so-called self-adjoint extension (SAE) parameter, defined as

$$‡_B = \frac{a_{add}}{a_{st}} \qquad (3.8)$$

$‡_B$ parameter is the same for all levels (with fixed orbital momentum $l$) and is real for bound states. From above considerations it is clear that we have three particular cases:

(i) If $a_{add} = 0 (‡_B = 0)$, we have only standard solutions.

(ii) If $a_{st} = 0 (‡_B = \infty)$, we keep only additional solutions.

(iii) If $‡_B \neq 0, \pm\infty$, solutions are neither pure standard nor pure additional. In this case SAE parameter becomes arbitrary one and it may be restricted only from some physical resonings.

## 4. What is new for Inverse square potential when we retain additional solutions?

Consider the following potential

$$V = -\frac{V_0}{r^2}, \qquad V_0 > 0 \qquad (4.1)$$

in the whole space. There is only one worthy case, namely $0 \leq P < 1/2$.

Now the wave function $R(r)$ for $E = 0$ has the form (2.9) in the whole space. It has a single zero, determined by

$$r_0 = \left(-\frac{B}{A}\right)^{1/2P} \qquad (4.2)$$

Therefore, the wave function has only one node and according to well-known theorem we have one bound state only. This result differs from that considered in any textbooks of quantum mechanics.

We can give very simple physical picture of how the additional solutions arise. For this purpose, let us rewrite the Schrodinger equation near the origin for attractive potential (4.11) in the form

$$R'' + \frac{2}{r}R' + 2m[E - V_{ac}(r)]R = 0 \qquad (4.3)$$

where

$$V_{ac} = \frac{P^2 - 1/4}{2mr^2} \qquad (4.4)$$

Consider the following possible cases:

i). If $P > 1/2$, then $V_{ac} > 0$ and it is repulsive centrifugal potential and as we saw, one has no additional solutions.

ii). If $0 \leq P < 1/2$, then $V_{ac} < 0$. Therefore, it becomes attractive and is called as quantum anti-centrifugal potential [4.7]. This potential has $R_{add}$ states, because the condition (2.4) is fulfilled in this case.

iii). If $P^2 < 0$, then $V_{ac}$ becomes strongly attractive and one has "falling to the center".

Therefore, the sign of the potential $V_{ac}$ determines whether we need additional solutions or not.

It was thought that potential (4.1) had no levels out of region of "falling to the center" (See e.g. [6,7]), but in [8-10] single level was found by complete SAE procedure, while the boundary condition and the range of parameter, like P, are questionable there. Here we'll show explicitly that this potential has exactly a single level, which depends on the SAE parameter $\tau$.

Let's take the Schrodinger equation for potential (4.1)

$$\frac{d^2R}{dr^2} + \frac{2}{r}\frac{dR}{dr} + \left(-k^2 - \frac{P^2 - 1/4}{r^2}\right)R = 0 \qquad (4.5)$$

where P is given by (2.10) and

$$k^2 = -2mE > 0; \quad (E < 0) \qquad (4.6)$$

One can reduce Eq.(4.5) to the equation for modified Bessel functions by substitutions

$$R(r) = \frac{f(r)}{\sqrt{r}}; \quad x = kr \qquad (4.7)$$

leading to the following equation

$$x^2 \frac{d^2 f(x)}{dx^2} + x \frac{df(x)}{dx} - \left(x^2 + P^2\right)f(x) = 0 \qquad (4.8)$$

This equation has 3 pairs of independent solutions[11]: $I_P(kr)$ and $I_{-P}(kr)$, $I_P(kr)$ and $e^{ifP}K_P(kr)$, $I_{-P}(kr)$ and $e^{ifP}K_P(kr)$, where $I_P(kr)$ and $K_P(kr)$ are Bessel and MacDonald modified functions [11], respectively. Consider these possibilities separately:

1) The pair $I_P(kr)$ and $I_{-P}(kr)$:

The general solution of (4.5) is

$$R = r^{-\frac{1}{2}}\left[AI_P(kr) + BI_{-P}(kr)\right] \qquad (4.9)$$

Consider the behaviour of this solution at small and large distances:

a) Small distances

In this case [11]

$$I_P(z) \underset{z \to 0}{\approx} \left(\frac{z}{2}\right)^P \frac{1}{\Gamma(P+1)} \tag{4.10}$$

Then it follows from (4.9) and (4.10) that

$$\lim_{r \to 0} R(r) \approx r^{-\frac{1}{2}} \left[ A\left(\frac{k}{2}\right)^P \frac{r^P}{\Gamma(P+1)} + B\left(\frac{k}{2}\right)^{-P} \frac{r^{-P}}{\Gamma(1-P)} \right] \tag{4.11}$$

From (2.9), (4.11), (3.9) and the definition (3.10) we obtain $\tau$,

$$\ddagger_B = \frac{B}{A} 2^P k^{-2P} \frac{\Gamma(1+P)}{\Gamma(1-P)} \tag{4.12}$$

b) Large distances

In this case [11]

$$I_P(z) \underset{z \to \infty}{\approx} \frac{e^z}{\sqrt{2fz}} \tag{4.13}$$

and

$$R(r) \underset{r \to \infty}{\approx} \frac{1}{\sqrt{2f}} \{A + B\} e^{kr} \tag{4.14}$$

Therefore, requiring vanishing of $R(r)$ at infinity, we have to take

$$B = -A \tag{4.15}$$

and from (4.12), (4.15) and (4.6) we obtain one real level (for fixed orbital $l$ momentum, satisfying (3.1)),

$$E = -\frac{2}{m} \left[ \frac{\Gamma(1+P)}{\Gamma(1-P)} \right]^{\frac{1}{P}} \left[ -\frac{1}{\ddagger_B} \right]^{\frac{1}{P}} ; \quad 0 < P < 1/2 \tag{4.16}$$

Eq. (4.16) is a new expression derived as a consequence of orthogonality condition in the framework of "pragmatic" approach.

In general, $\tau$ is a free parameter but some physical requirements may restrict its magnitude. For example, reality of energy in (4.16) restricts $\tau$ parameter to be negative $\ddagger < 0$. Note also that, as it is clear from the derivation of (4.16), this level disappears for standard quantum mechanics ($\ddagger = 0$) and $\ddagger = \pm\infty$, and for these values scale invariance is restored.

To obtain corresponding wave function, take into account a well-known relation [11]

$$K_P(z) = \frac{f}{2\sin Pf} [I_{-P}(z) - I_P(z)] \tag{4.17}$$

Then the wave function corresponding to the level (4.16) is

$$R = -A\frac{2}{f} r^{-\frac{1}{2}} \sin Pf \cdot K_P(kr) \tag{4.18}$$

Because of exponential damping

$$K_P(z) \underset{z\to\infty}{\approx} \sqrt{\frac{f}{2z}} e^{-z} \tag{4.19}$$

the function (4.18) corresponds to the bound state. It is also known that $K_P(z)$ function has no zeroes for real P $(0 < P < 1/2)$ and therefore (4.16) corresponds to a single bound state. Moreover, wave function (4.18) satisfies the fundamental condition (4.16) for $0 < P < 1/2$.

2) The pair $I_P(kr)$ and $e^{ifP} K_P(kr)$;

The general solution of (4.5) is

$$R = r^{-\frac{1}{2}} \left[ AI_P(kr) + Be^{ifP} K_P(kr) \right] \tag{4.20}$$

At large distances

$$\lim_{r\to\infty} R(r) \approx \frac{1}{\sqrt{2f}} \left( Ae^{kr} + Be^{ifP} f e^{-kr} \right) \approx A \frac{e^{kr}}{\sqrt{2f}} \tag{4.21}$$

Therefore, we have no bound states.

The same follows for pair $I_{-P}(kr)$ and $e^{ifP} K_P(kr)$. Thus, only pair $I_P(kr)$ and $I_{-P}(kr)$ has a single bound state.

Noting that the consideration of all possible pairs of solution is, in general, necessary, because there is no guide principle, by which one can guess which pair must be considered.

### V. Orthogonality requirement in the scattering problems

Let us return to pure inverse square potential (4.1) again.

For scattering problem the general solution of Schrodinger equation with potential (4.1) is

$$R(r) = \sqrt{k/r} \left\{ A(k) J_P(kr) + B(k) J_{-P}(kr) \right\}; \quad k^2 = 2mE, \quad E > 0 \tag{5.1}$$

The orthogonality condition, that must obey the radial function, now has a form

$$I = \int_0^\infty r^2 dr R_{k'}^*(r) R_k(r) = 2f u(k' - k) \tag{5.2}$$

After substitution the Eq.(5.1) into (5.2) we meet integrals like

$$\int_0^\infty J_P(k'r) J_P(kr) r\,dr = \frac{\mathrm{u}(k'-k)}{\sqrt{k'k}}, \tag{5.3}$$

$$\int_0^\infty J_P(k'r) J_{-P}(kr) r\,dr = \frac{\mathrm{u}(k'-k)}{\sqrt{k'k}} \cos f P + \frac{2 \sin f P}{f(k^2 - k'^2)} \left(\frac{k}{k'}\right)^P, \tag{5.4}$$

Using them, we derive

$$I = \left\{ AA^* + BB^* + (AB^* + A^*B) \cos f P \right\} \mathrm{u}(k'-k) + \frac{2 \sin f P}{f(k^2 - k'^2)} \left\{ \left(\frac{k}{k'}\right)^P B^*(k') A(k) - \left(\frac{k}{k'}\right)^{-P} A^*(k') B(k) \right\}$$

(5.5)

Comparing (5.2) and (5.5), we conclude that in order to satisfy the Eq.(5.2) the vanishing of last bracket in (5.5) is necessary, i.e.

$$\frac{B^*(k')}{A^*(k')} (k')^{-2P} = \frac{B(k)}{A(k)} (k)^{-2P} \equiv \ddagger_S \tag{5.6}$$

where the SAE parameter $\ddagger_S$ is introduced for scattering processes. This form shows that both side of Eq.(5.6) is independent of $k$ and, moreover, $\ddagger_S$ is a real number. Taking into account (5.6) in the first line of (5.5), one can exclude the parameter $B$ in favour of $A$ and suppose $k' = k$, allowing by the overall delta function factor, we can equate the paranthesis to $2f$ and archive to fulfilment of orthogonality condición (5.2), require

$$AA^* \left\{ \ddagger_S^2 k^{4P} + 2 \ddagger_S k^{2P} \cos f P + 1 \right\} = 2f \tag{5.7}$$

It is remarkable to note that above relations can be derived more transparently by means of the Wronskian method. Using the radial equation (2.1) for both functions $R_k$ and $R_{k'}^*$ by obvious manipulations it is easy to show that

$$\lim_{r \to \infty} \int_0^r R_{k'}^*(r) R_k(r) r^2 dr = \lim_{r \to \infty} \frac{1}{k'^2 - k^2} \left[ u_{k'}^* \frac{du_k}{dr} - u_k \frac{du_{k'}^*}{dr} \right]_0^r \tag{5.8}$$

Where, for convenience, we have temporarily introduced the notation

$$u_j(r) = r R_j(r) = \sqrt{k_j r} \left\{ A(k_j) J_P(k_j r) + B(k_j) J_{-P}(k_j r) \right\} \tag{5.9}$$

So, the orthogonality integral is expressed by the Wronskian of the reduced radial wave functions.

From the asymptotic of the Bessel functions at the origin

$$J_P(z) \underset{z \to 0}{\approx} \left(\frac{z}{2}\right)^P \frac{1}{\Gamma(P+1)} \tag{5.10}$$

we have

$$u_k(r) \approx a_{st}(k) r^{1/2+P} + a_{add}(k) r^{1/2-P}, \quad \text{at } r \to 0 \tag{5.11}$$

where

$$a_{st}(k) = \frac{k^{1/2+P}}{2^P \Gamma(1+P)} A; \qquad a_{add}(k) = \frac{k^{1/2-P}}{2^{-P} \Gamma(1-P)} B \tag{5.12}$$

Then Eq.(5.8) gives at the lower boundary

$$I\big|_0 = \frac{2P}{k'^2 - k^2} \{ a_{st}(k) a_{add}^*(k') - a_{st}^*(k') a_{add}(k) \} \tag{5.13}$$

which is very like to the case of discrete states. For orthogonality one must require

$$\frac{a_{add}^*(k')}{a_{st}^*(k')} = \frac{a_{add}(k)}{a_{st}(k)} \tag{5.14}$$

It generalizes the definition of SAE parameter for discrete case (3.9) and we see that $\ddagger_B = \ddagger_S$. As regards of the relation (5.2), it follows from uper boundary of the same integral (5.8).

## VI. Partial Scattering Amplitude and Phase shift

The behavior of Bessel function at large distances is

$$J_P(kr) \approx \sqrt{\frac{2}{f kr}} \cos\left( kr - \frac{Pf}{2} - \frac{f}{4} \right) \quad _{r\to\infty} \tag{6.1}$$

It gives the following asymptotic of radial function

$$\lim_{r\to\infty} R \approx \frac{1}{\sqrt{2f}\, r} e^{ikr} \left[ A e^{-i(P+1/2)\frac{f}{2}} + B e^{i(P-1/2)\frac{f}{2}} \right] + \frac{1}{\sqrt{2f}\, r} e^{-ikr} \left[ A e^{i(P+1/2)\frac{f}{2}} + B e^{-i(P-1/2)\frac{f}{2}} \right] \tag{6.2}$$

On the other hand, using the definition of scattering phase $\mathsf{u}_l$

$$\lim_{r\to\infty} R \approx \frac{2}{r} \sin\left( kr - \frac{lf}{2} + \mathsf{u}_l \right) \tag{6.3}$$

one can write

$$\lim_{r\to\infty} R \approx \frac{1}{ir} \left\{ e^{ikr} e^{i(\mathsf{u}_l - f l/2)} - e^{-ikr} e^{i(-\mathsf{u}_l + f l/2)} \right\} \tag{6.4}$$

Comparison with Eq.(5.2) gives

$$\frac{1}{\sqrt{2f}}\left[Ae^{-i(P+1/2)\frac{f}{2}}+Be^{i(P-1/2)\frac{f}{2}}\right]=\frac{1}{i}e^{i(u_l-fl/2)} \tag{6.5}$$

$$\frac{1}{\sqrt{2f}}\left[Ae^{i(P+1/2)\frac{f}{2}}+Be^{-i(P-1/2)\frac{f}{2}}\right]=ie^{-i(u_l-fl/2)} \tag{6.6}$$

From these relations we derive the partial scatering amplitude

$$S_l=e^{2iu_l}=\frac{Ae^{-i(P+1/2)\frac{f}{2}}+Be^{i(P-1/2)\frac{f}{2}}}{Ae^{i(P+1/2)\frac{f}{2}}+Be^{-i(P-1/2)\frac{f}{2}}}e^{if(l+1)} \tag{6.7}$$

and finaly, by using Eq. (5.6) we get

$$S_l=e^{2i(l+1/2-P)\frac{f}{2}}\frac{1+\ddagger_s k^{2P}e^{ifP}}{1+\ddagger_s k^{2P}e^{-ifP}} \tag{6.8}$$

Remark that here the fraction is a new factor, which appears owing to SAE procedure.

Let us show that the Eq. (6.7) gives correct physical results.

(a) when $\ddagger_s=0$ (or $B=0$) the standard result follows

$$S_l=S_l^{st}=\exp\left[2i(l+1/2-P)\frac{f}{2}\right] \tag{6.9}$$

(b) when $\ddagger_s=\pm\infty$ (or $A=0$), it follows additional partial amplitude

$$S_l^{add}=e^{2i(l+1/2+P)\frac{f}{2}}, \tag{6.10}$$

which can be derived from previous result by changing $P\to-P$.

(c) Let us now derive the additional bound state energy as a pole of (6.8). It is evident that this pole must be emerged from the additional multiplier of (6.8) as a consequence of SAE. Consider this fraction

$$Q=\frac{1+\ddagger_s\frac{\Gamma(1-P)}{\Gamma(1+P)}k^{2P}e^{ifP}}{1+\ddagger_s\frac{\Gamma(1-P)}{\Gamma(1+P)}k^{2P}e^{-ifP}} \tag{6.11}$$

and perform needed change for transition to bound states, namely $k\to i|$ or $k^2=-|^2=2mE$, $(E<0)$. In this case (6.8) has a pole at

$$(-|^2)^P\ddagger_s e^{-ifP}=-1 \tag{6.12}$$

or at

$$E = -\frac{2}{m}\left[\frac{\Gamma(1+P)}{\Gamma(1+P)}\right]^{1/P}\left[-\frac{1}{\ddagger_S}\right]^{1/P}, \quad 0 < P < 1/2, \tag{6.13}$$

which coincides with (4.2) because $\ddagger_S = \ddagger_B$

Now we want rewrite a scattering phase in different form by using following relation

$$e^{2i\tan^{-1} z} = \frac{1+iz}{1-iz} \tag{6.14}$$

then (6.7) can be rewritten as

$$S_l = e^{2i\tilde{u}_l}, \tag{6.15}$$

where

$$\tilde{u}_l = u_{st} + u_{SAE} = [l+1/2-P]\frac{f}{2} + \tan^{-1}\frac{\ddagger_S k^{2P}\sin f P}{1+\ddagger_S k^{2P}\cos f P} \tag{6.16}$$

Here the second term is new and gives rise from SAE procedure

$$u_{SAE} = \tan^{-1} X, \qquad X = \frac{\ddagger_S k^{2P}\sin f P}{1+\ddagger_S k^{2P}\cos f P} \tag{6.17}$$

Let us make some comments:

(i) This part of phase shift depends on energy $(k^2 = 2mE)$, therefore it violates the scale invariance, which may be restored only for three values of SAE parameter, $\ddagger_S = 0$ and $\ddagger_S = \pm\infty$.

(ii) We considered from the beginning attractive potential, for which $u_{st} > 0$. But it seems from (6.16) that the full phase the shift may become negative, or potential acquires repulsive character. There appears two alternatives: restrict the SAE parameter so that potential remains attractive -- $\tilde{u}_l > 0$ or allow the alteration of nature of potential by SAE procedure.

## VII. CONCLUSIONS

In this article we have considered a pragmatic approach for a self-adjoint extension in the Schroedinger equation in case of an inverse square potential in both disctrete as well as scattering states. Our consideration concerns to the three dimentional quantum mechanics. As an explicit example we considered a pure $r^{-2}$ attractive potential. The following new results are derived:

(i) We find a single bound state depended on SAE parameter;
(ii) Generalized the method to continuous spectrum (scattering processes);

(iii) We find a modified partial wave amplitude after SAE, which consists an extra factor owing to the extension procedure, that has a single pole for bound state;

In conclusion let us note that Hamiltonians of physically significant quantum-mechanical problems manifests in such $r^{-2}$ behavior. Hamiltonians with inverse square behavior appear in many systems and they have sufficiently rich physical and mathematical structures Examples of such systems are: conformal quantum mechanics [12], Aharonov-Bohm effect [13], Dirac monopoles [14], valence electron model for hydrogen-like atoms in quantum mechanics [15], the theory of black -holes [16], Calogero model [17], etc. At small distances $r^{-2}$ like potentials have singular solutions together with regular ones. As a rule such solutions are ignored from consideration, but by our opinion this action is not always reasonable and legitimate. Our investigation above substantiates evident usefulness of used methods. Therefore, we will return to these problems in subsequent publications.